\documentclass[amsmath,amssymb,reprint,twocolumn, prb, superscriptaddress]{revtex4-1}

\usepackage[utf8]{inputenc}
\usepackage[T1]{fontenc}

\usepackage[english]{babel}

\usepackage{amsmath,amssymb,amsfonts}
\usepackage{amstext, mathrsfs, textcomp}
\usepackage{mathptmx}
\usepackage{bigints}  
\usepackage{nicefrac}
\usepackage{multirow}
\usepackage{dcolumn}
\usepackage{bm}

\usepackage{subfigure}
\usepackage{graphicx}
\usepackage{xcolor}
\usepackage[export]{adjustbox}

\usepackage[colorlinks=true, citecolor=blue, urlcolor=blue, linkcolor=blue]{hyperref}

\usepackage[final]{changes}

\newcommand{\im}{\text{i}}

\usepackage{bm} 
\usepackage{fixmath} 
\usepackage{scalerel}
\newlength\bshft
\def\fakebold#1{\ThisStyle{\ooalign{$\SavedStyle#1$\cr%
  \kern-\bshft$\SavedStyle#1$\cr%
  \kern\bshft$\SavedStyle#1$}}}

\renewcommand{\Im}{\text{Im}}

\graphicspath{{figures/}}

\begin{document}

\title{Polarizabilities of complex individual dielectric or plasmonic nanostructures}

\author{\firstname{Adelin} \surname{Patoux}}
\affiliation{CEMES, Universit\'e de Toulouse, CNRS, Toulouse, France}
\affiliation{LAAS, Universit\'e de Toulouse, CNRS, Toulouse, France}
\affiliation{AIRBUS DEFENCE AND SPACE SAS, Toulouse, France}

\author{\firstname{Cl\'ement} \surname{Majorel}}
\affiliation{CEMES, Universit\'e de Toulouse, CNRS, Toulouse, France}

\author{\firstname{Peter R.} \surname{Wiecha}}
\email[e-mail~: ]{p.wiecha@soton.ac.uk}
\affiliation{CEMES, Universit\'e de Toulouse, CNRS, Toulouse, France}
\affiliation{Physics and Astronomy, Faculty of Engineering and Physical Sciences, University of Southampton, Southampton, UK}

\author{\firstname{Aur\'elien} \surname{Cuche}}
\affiliation{CEMES, Universit\'e de Toulouse, CNRS, Toulouse, France}

\author{\firstname{Otto L.} \surname{Muskens}}
\affiliation{Physics and Astronomy, Faculty of Engineering and Physical Sciences, University of Southampton, Southampton, UK}

\author{\firstname{Christian} \surname{Girard}}
\affiliation{CEMES, Universit\'e de Toulouse, CNRS, Toulouse, France}

\author{\firstname{Arnaud} \surname{Arbouet}}
\email[e-mail~: ]{arnaud.arbouet@cemes.fr}
\affiliation{CEMES, Universit\'e de Toulouse, CNRS, Toulouse, France}

\begin{abstract}
When the sizes of photonic nanoparticles are much smaller than the excitation wavelength, their optical response can be efficiently described with a series of polarizability tensors.
Here, we propose a universal method to extract the different components of
the response tensors associated with small plasmonic or dielectric particles.
We demonstrate that the optical response can be faithfully approximated, as long as the effective dipole is not induced by retardation effects, hence do not depend on the phase of the illumination. 
We show that the conventional approximation breaks down for a phase-driven dipolar response, such as optical magnetic resonances in dielectric nanostructures. 
To describe such retardation induced dipole resonances in intermediate-size dielectric nanostructures, we introduce ``pseudo-polarizabilities'' including first-order phase effects, which we demonstrate at the example of magnetic dipole resonances in dielectric spheres and ellipsoids.
Our method paves the way for fast simulations of large and inhomogeneous meta-surfaces.
\end{abstract}

\maketitle 

\section{Introduction}
In a multitude of topical areas in contemporary physics and chemistry, the concept of the polarizability has proven to be extremely useful.
In particular, in the physics of gases and surfaces, the dynamic polarizability tensor of molecules appears explicitly in the description, for example, of the Van der Waals dispersion energy, or in the description of the Raman scattering process.\cite{girardEffectivePolarizabilityMolecule1987, girardEffectivePolarizabilityTwo1987, buckinghamPolarizabilityPairInteracting1956, a.d.buckinghamPermanentInducedMolecular1967, barronRayleighRamanOptical1975, buckinghamIntermolecularForces1975}
During the 1970s, A. D. Buckingham wrote a founding article on this subject in which an exhaustive list of linear polarizabilities is proposed.\cite{buckinghamPolarizabilityHyperpolarizabilityDiscussion1979}
Although this work was restricted to atomic and molecular systems, it represents a valuable stand of the various possible contributions as well as their ranking, in terms of electric and magnetic multipolar polarizabilities.

The theoretical study of the linear optical response of small metallic or dielectric particles has also been extensively investigated in the last decades. 
In particular, in the context of plasmonics the concept of polarizability is often applied to the description of plasmon 
spectra of sub-wavelength sized noble metal particles.\cite{evlyukhinOpticalResponseFeatures2010, sersicMagnetoelectricPointScattering2011, bowenUsingDiscreteDipole2012, arangoPolarizabilityTensorRetrieval2013, bernalarangoUnderpinningHybridizationIntuition2014, asadchyBianisotropicMetasurfacesPhysics2018, bertrandGlobalPolarizabilityMatrix2019}
In many situations, single metal particles can be schematized by a sphere of radius $a$, in which case their optical response can be described by a scalar, frequency-dependent polarizability $\alpha(\omega_{0})$. 
Then, the polarizability tensor is diagonal and all tensor elements are identical.
In cgs units, it reads:\cite{draineDiscreteDipoleApproximationIts1988}
\begin{equation}
\alpha_{ij}(\omega_{0})=a^{3}\left(\frac{\epsilon_{2}(\omega_{0})-\epsilon_{1}}
{\epsilon_{2}(\omega_{0})+2\epsilon_{1}}\right)
\; ,
\label{ALPHA1}
\end{equation}
where $\epsilon_{1}$ (respectively $\epsilon_{2}$) is the dielectric constant of the medium (respectively the nano-sphere). 
From relation (\ref{ALPHA1}), we  can extract the extinction spectrum via the imaginary part of $\alpha(\omega_{0})$.
Consequently, the extinction spectra of a sample containing a large number \(N\) of such non--interacting nanoparticles $\alpha_i(\omega_{0})$ is given by:\cite{draineDiscreteDipoleApproximationIts1988, girardShapingManipulationLight2008} 
\begin{equation}
I_{\text{ext}}(\lambda_{0})=\frac{8 \pi^{2}}{n_1 \lambda_{0}} \sum\limits_i^N \Im \Big( \alpha_i(\omega_{0}) \Big)
\; ,
\label{EXTINC}
\end{equation}
where $\lambda_{0}$ represents the incident wavelength, \(n_1\) the refractive index of the environment, and ``\(\Im\)'' the imaginary part.

The sphere represents the highest symmetry, belonging to the {\it isotropic} symmetry group.
As stated above, in this case, all the diagonal elements of the polarizability are identical, and the system displays a scalar response defined by $\alpha_{ij}=\alpha \delta_{ij}$ (see equations (\ref{ALPHA1}) and (\ref{EXTINC})).
When transforming the sphere into an ellipsoid of symmetry group $D_{\infty h}$, the polarizability
must be defined with two independent components,\cite{morozDepolarizationFieldSpheroidal2009} and for even lower symmetry, 
all components $\alpha_{ij}$ of the polarizability tensor must be calculated.
This situation corresponds to high anisotropy induced by a complex shape of particles (or nano-cavities).
Note that other kinds of anisotropy can come from the 
intrinsic anisotropy of the dielectric constant of the particle but also from the surface of another object.\cite{sihvolaElectromagneticWavesChiral1994}
In the latter case, the concept of effective polarizability is generally introduced,
and the final symmetry of the particle is dressed by the symmetry of the surface
(i.e. $D_{\infty h}$, for a perfectly planar surface). 

As illustrated by these examples, the design of nanostructure polarizabilities starts with the conception of a reference geometry by intuitive considerations.
Such an approach, however, is limited to rather simple problems. In case of complex structures or complicated phenomena, the intuitive method often fails, as unexpected effects such as polarization conversion occur in the polarizability tensors.
In this work we propose a numerical method to extract the polarizability tensors for complex shaped metallic and dielectric nanostructures through a volume discretization technique, which uses the concept of a {\it generalized propagator}.
Furthermore, in order to faithfully describe also magnetic optical effects in dielectric nanostructures, where the conventional dipolar polarizability approximation fails, we introduce ``pseudo-polarizabilities'' that include phase-induced magnetic dipole resonances, similar to some homogenization approaches for metamaterials,\cite{aluFirstprinciplesHomogenizationTheory2011, ciattoniNonlocalHomogenizationTheory2015} but at the level of a single, isolated structure.
Our pseudo-polarizabilities might then be used to construct aperiodic or random metasurface-like assemblies without periodicity.


\section{A Generalized Electromagnetic Propagator for Arbitrary Shaped Particles or Cavities}

The concept of the {\it generalized electric field propagator} previously described in reference~\onlinecite{martinGeneralizedFieldPropagator1995} can be easily extended to the general case of meta-systems displaying both an electric and a magnetic linear response. In this case, the source zone as depicted in figure~\ref{fig:zone} is characterized by the following susceptibility tensor, where \(\fakebold{\mathbb{I}}\) is the identity tensor:

\begin{figure}[t]
\centering\includegraphics[width=\columnwidth]{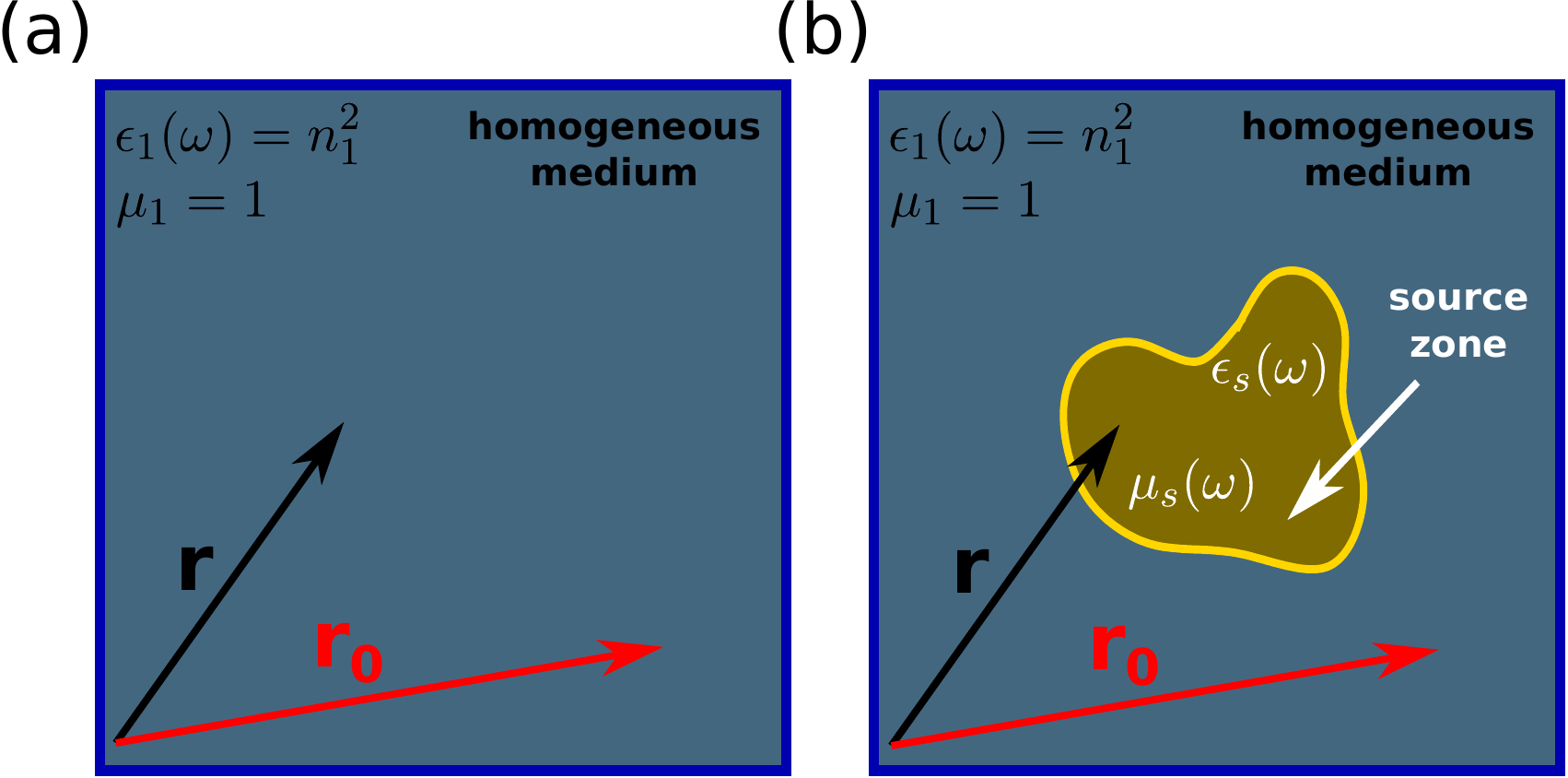}
\caption{
Sketch used to implement the concept of generalized electromagnetic propagator.
(a) transparent reference medium with $\epsilon_{1}(\omega_0)=n_{1}^{2}$ and $\mu_{1}=1$;
(b) material system of arbitrary shape, also called the
{\it source zone}, embedded in the reference medium (permittivity $\epsilon_{s}(\omega_0)$ 
and permeability $\mu_{s}(\omega_0)$).
}
\label{fig:zone}
\end{figure}

\begin{equation}
\bm{\chi}(\omega_0)=
\left( \begin{matrix}
\chi_{e}(\omega_0)\, \fakebold{\mathbb{I}} & 0 \\[12pt]
0 & \chi_{m}(\omega_0) \fakebold{\mathbb{I}}
\end{matrix}\right)
\; ;
\label{CHI}
\end{equation}
where $\chi_{e}(\omega_0)$ and $\chi_{m}(\omega_0)$ are related to the permittivity $\epsilon_{s}(\omega_0)$, respectively the permeability $\mu_{s}(\omega_0)$ of the source zone:
\begin{equation}
\chi_{e}(\omega_0)=\frac{\epsilon_{s}(\omega_0)-\epsilon_{1}(\omega_0)}{4\pi}
\; ;
\label{CHI-E}
\end{equation}
and
\begin{equation}
\chi_{m}(\omega_0)=\frac{\mu_{s}(\omega_0)-\mu_{1}(\omega_0)}{4\pi}
\; .
\label{CHI-M}
\end{equation}
Introducing two super vectors 
$\mathbf{F}_{0}(\mathbf{r},\omega_0) = (\mathbf{E}_{0}(\mathbf{r},\omega_0),\, \mathbf{H}_{0}(\mathbf{r},\omega_0))$
and $\mathbf{F}(\mathbf{r},\omega_0) = (\mathbf{E}(\mathbf{r},\omega_0),\, \mathbf{H}(\mathbf{r},\omega_0))$ 
(where $\mathbf{E}$ and $\mathbf{H}$ refer to electric and magnetic fields, respectively)
to describe the incident and total electromagnetic
fields, we can define a unique $(6\times 6)$ dyadic tensor
$\protect\fakebold{\mathbb{K}}(\mathbf{r},\mathbf{r}',\omega_0)$
operating in the volume $V$
of the source zone and 
establishing the link between $\mathbf{F}_{0}(\mathbf{r},\omega_0)$ and $\mathbf{F}(\mathbf{r},\omega_0)$:
\begin{equation}
\mathbf{F}(\mathbf{r},\omega_0)=\int_{V}\protect\fakebold{\mathbb{K}}(\mathbf{r},\mathbf{r}',\omega_0)
\cdot\mathbf{F}_{0}(\mathbf{r}',\omega_0)\mathrm{d}\mathbf{r}'\; .
\label{EQ-INT-K}
\end{equation}
Actually, the $(6\times 6)$ superpropagator $\protect\fakebold{\mathbb{K}}(\mathbf{r},\mathbf{r}',\omega_0)$
is composed of four mixed $(3\times 3)$ dyadic tensors:
\begin{equation}
\protect\fakebold{\mathbb{K}}(\mathbf{r},\,\mathbf{r}',\omega_0)=
\left( \begin{matrix}
K^{EE}(\mathbf{r},\mathbf{r}',\omega_0) & K^{EH}(\mathbf{r},\mathbf{r}',\omega_0) \\[12pt]
K^{HE}(\mathbf{r},\mathbf{r}',\omega_0) & K^{HH}(\mathbf{r},\mathbf{r}',\omega_0)
\end{matrix}\right)
\label{K-KMIX}
\end{equation} 
in which the first one, $K^{EE}(\mathbf{r},\mathbf{r}',\omega_0)$ that describes the {\it electric--electric field
couplings} was introduced in the early beginning of {\it near--field optics} \cite{martinGeneralizedFieldPropagator1995}.
The three other contributions, i.e. $K^{EH}(\mathbf{r},\mathbf{r}',\omega_0)$,
$K^{HE}(\mathbf{r},\mathbf{r}',\omega_0)$ and $K^{HH}(\mathbf{r},\mathbf{r}',\omega_0)$, account for coupling with
the magnetic field. All these propagators are related to the corresponding {\it mixed field--susceptibilities}
$S^{EE}$, $S^{EH}$, $S^{HE}$, and $S^{HH}$, 
\cite{sersicMagnetoelectricPointScattering2011, wiechaDecayRateMagnetic2018} associated with the {\it source zone}:
\begin{equation}
\begin{aligned}
&\mathbf{K}^{EE}(\mathbf{r},\mathbf{r}',\omega_0)=\delta(\mathbf{r}-\mathbf{r}')\protect
\fakebold{\mathbb{I}}+\chi_{e}(\omega_0)\cdot S^{EE}(\mathbf{r},\mathbf{r}',\omega_0)\\[2pt]
&\mathbf{K}^{EH}(\mathbf{r},\mathbf{r}',\omega_0)=\chi_{m}(\omega_0)\cdot S^{EH}(\mathbf{r},\mathbf{r}',\omega_0)\\[2pt]
&\mathbf{K}^{HE}(\mathbf{r},\mathbf{r}',\omega_0)=\chi_{e}(\omega_0)\cdot S^{HE}(\mathbf{r},\mathbf{r}',\omega_0)\\[2pt]
&\mathbf{K}^{HH}(\mathbf{r},\mathbf{r}',\omega_0)=\delta(\mathbf{r}-\mathbf{r}')\protect
\fakebold{\mathbb{I}}+\chi_{m}(\omega_0)\cdot S^{HH}(\mathbf{r},\mathbf{r}',\omega_0)
\end{aligned}
\label{MIXED-S}
\end{equation}
As explained in references~\onlinecite{martinGeneralizedFieldPropagator1995} and~\onlinecite{wiechaDecayRateMagnetic2018}, these dyadic tensors can be
numerically computed by performing a volume discretization of the {\it source zone}
together with a {\it Dyson sequence procedure}\cite{martinGeneralizedFieldPropagator1995} or other numerical inversion techniques, to extract the various field-susceptibilities
in the {\it source zone}.

\section{Extraction of polarizabilities of small nanostructures}

The volume discretization of the {\it source region} leads to a mesh of $N$ identical elementary volumes $\Delta v$.
Such a procedure converts integrals over the source volume $V$ into discrete summations.
In reference \cite{girardShapingManipulationLight2008} we have gathered the expressions of the discretization volume elements \(\Delta v\) for both cubic and hexagonal compact discretization grids together with the corresponding Green's function renormalization terms.
The electric polarization at the {\it i}th cell in the source region can be written as follows:
\begin{equation}
\label{Pola_elec}
\begin{aligned}
\mathcal{P}(\mathbf{r}_{i},\omega_{0})= 
\Delta v^{2} & \chi_e(\omega_0) 
\\
\times \sum_{j=1}^{N} \Big( & \mathbf{K}^{EE}(\mathbf{r}_{i},\mathbf{r}_{j},\omega_{0})
\cdot\mathbf{E}_{0}(\mathbf{r}_{j},\omega_{0})
\\
& + \mathbf{K}^{EH}(\mathbf{r}_{i},\mathbf{r}_{j},\omega_{0})
\cdot\mathbf{H}_{0}(\mathbf{r}_{j},\omega_{0})\Big) \, .
\end{aligned}
\end{equation} 
\noindent
Concerning the magnetic polarization $\mathcal{M}(\mathbf{r}_{i},\omega_{0})$ induced in the {\it source region}, it
may be split into two contributions related to $\chi_{e}(\omega_{0})$ and $\chi_{m}(\omega_{0})$, respectively:
\begin{equation}
\label{Pola_mag}
\mathcal{M}(\mathbf{r}_{i},\omega_{0})=\mathcal{M}_{e}(\mathbf{r}_{i},\omega_{0})
+\mathcal{M}_{m}(\mathbf{r}_{i},\omega_{0})
\end{equation}
with
\begin{equation}
\label{Pola_mage}
\begin{aligned}
\mathcal{M}_{e}(\mathbf{r}_{i},\omega_{0}) = &-\dfrac{ik_{0}}{2}\Delta v^{2} \chi_{e}(\omega_{0})
\\
\times \sum_{j=1}^{N} 
& \Big(\mathbf{r}_{i}\wedge\mathbf{K}^{EE}
(\mathbf{r}_{i},\mathbf{r}_{j},\omega_{0})\cdot\mathbf{E}_{0}(\mathbf{r}_{j},\omega_{0})
\\
& + \mathbf{r}_{i}\wedge\mathbf{K}^{EH}(\mathbf{r}_{i},\mathbf{r}_{j},\omega_{0})\cdot\mathbf{H}_{0}(\mathbf{r}_{j},\omega_{0})\Big)
\end{aligned}
\end{equation}
and
\begin{equation}
\label{Pola_MAGm}
\begin{aligned}
\mathcal{M}_{m}(\mathbf{r}_{i},\omega_{0}) = & \Delta v^{2}\chi_{m}(\omega_{0})
\\
\times \sum_{j=1}^{N} &
\Big(\mathbf{K}^{HE}(\mathbf{r}_{i},\mathbf{r}_{j},\omega_{0})\cdot\mathbf{E}_{0}(\mathbf{r}_{j},\omega_{0})
\\
& + \mathbf{K}^{HH}(\mathbf{r}_{i},\mathbf{r}_{j},\omega_{0})\cdot\mathbf{H}_{0}(\mathbf{r}_{j},\omega_{0})\Big)
\; ,
\end{aligned}
\end{equation}
where the first contribution in Eq.~(\ref{Pola_MAGm}), proportional to $k_{0}$ = $\omega_{0}/c$, originates from polarization
vortices induced by phase changes inside the source region. 
These magnetic polarization effects have been extensively 
studied recently in the case of high index dielectric nano-structures.\cite{evlyukhinOpticalResponseFeatures2010, kuznetsovMagneticLight2012, albellaLowLossElectricMagnetic2013, albellaElectricMagneticField2014, deckerResonantDielectricNanostructures2016, barredaRecentAdvancesHigh2019}
Note that the choice of the center of the coordinate system is important, as it has an impact on the magnetic polarization \(\mathcal{M}_e\).
Usually, it is convenient to use the center of mass $\mathbf{r}_{c}$ of the nanostructure\cite{evlyukhinMultipoleLightScattering2011} and we will adopt this choice for the following examples where we set $\mathbf{r}_{c}$ as the center of the coordinate system.

The total electric polarization $\mathcal{P}(\omega_{0})$ (respectively magnetic polarization $\mathcal{M}(\omega_{0})$) is obtained by adding the local electric polarizations Eq.~(\ref{Pola_elec}) (respectively the magnetic polarizations Eq.~(\ref{Pola_mag})) of all the elementary cells of the volume discretization.
These polarizations are related to the super vector $\mathbf{F}_{0}(\mathbf{r}_{c},\omega)$ at the center of mass $\mathbf{r}_{c}$ of the nanostructure by the $(6\times6)$ {\it super polarizability} $\bm{\alpha}(\omega_{0})$:

\begin{figure}[t]
	\centering\includegraphics[width=0.7\columnwidth]{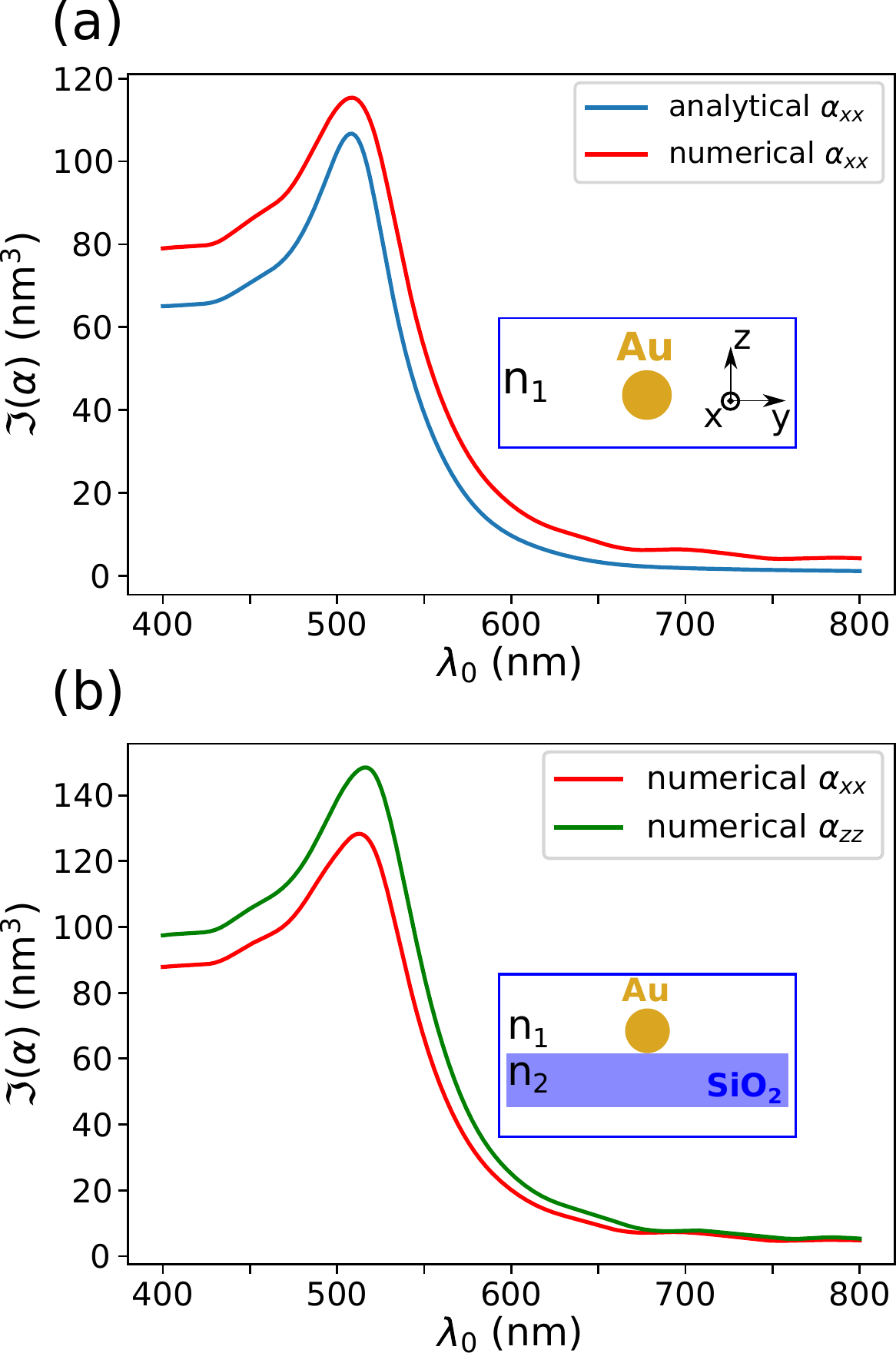}
	\caption{
	Spectral variation of the imaginary part of the dipolar polarizability of a spherical gold particle of radius 5nm. 
	(a) nano-sphere suspended in vacuum ($n_{1}$=1). Comparison of the analytical Clausius-Mossotti polarizability (blue) with the numerical calculation (red). 
	(b) numerically calculated in-plane (red) and out-of-plane (green) polarizability tensor elements for a nano-sphere deposited on a silica substrate ($n_{2}$=1.48). 
	Insets show sketches of the particle environment. 
	For the numerical calculations, we discretized the spheres using 6337 identical mesh cells on a hexagonal compact grid.
	}
	\label{fig:sphere_polarizability}
\end{figure}

\begin{equation}
\left( \begin{matrix}
\mathcal{P}(\omega_{0}) \\[12pt]
\mathcal{M}(\omega_{0})
\end{matrix}\right)=
\overbrace{
	\left( \begin{matrix}
	\alpha^{EE}(\omega_{0}) & \alpha^{EH}(\omega_{0}) \\[12pt]
	\alpha^{HE}(\omega_{0}) & \alpha^{HH}(\omega_{0}) 
	\end{matrix}\right)
}^{\bm{\alpha}(\omega_{0})}
\cdot
\overbrace{
\left( \begin{matrix}
\mathbf{E}_{0}(\mathbf{r}_{c},\omega_{0}) \\[12pt]
\mathbf{H}_{0}(\mathbf{r}_{c},\omega_{0})
\end{matrix}\right)}^{\mathbf{F}_0(\mathbf{r}_c, \omega_{0})}
\end{equation}
where the polarizabilities $\alpha^{EE}(\omega_{0})$, $\alpha^{EH}(\omega_{0})$, $\alpha^{HE}(\omega_{0})$ and $\alpha^{HH}(\omega_{0}) $ are four (3$\times$3) dyadic tensors, defined by
\begin{subequations}\label{eq:different_alphas}
\begin{equation}\label{alpha_EE}
\alpha^{EE}(\omega_{0})=\Delta v^{2}\chi_{e}(\omega_{0})\sum_{i,j}^{N}\mathbf{K}^{EE}(\mathbf{r}_{i},\mathbf{r}_{j},\omega_{0})e^{\im\mathbf{k}\cdot\mathbf{r}_{j}}
\end{equation}
\begin{equation}
\alpha^{EH}(\omega_{0})=\Delta v^{2}\chi_{e}(\omega_{0})\sum_{i,j}^{N}\mathbf{K}^{EH}(\mathbf{r}_{i},\mathbf{r}_{j},\omega_{0})e^{\im\mathbf{k}\cdot\mathbf{r}_{j}}
\end{equation}
\begin{equation}
\begin{aligned}
\alpha^{HE}&(\omega_{0})=\Delta v^{2}\sum_{i,j}^{N}\Big\lbrace \chi_{m}(\omega_{0})\mathbf{K}^{HE}(\mathbf{r}_{i},\mathbf{r}_{j},\omega_{0})\\[1pt]
&-\dfrac{ik_{0}}{2}\chi_{e}(\omega_{0})\mathbf{r}_{i}\wedge\mathbf{K}^{EE}(\mathbf{r}_{i},\mathbf{r}_{j},\omega_{0})\Big\rbrace e^{\im\mathbf{k}\cdot\mathbf{r}_{j}}\\[2pt]
\end{aligned}\label{eq:alpha_HE}
\end{equation}
\begin{equation}
\begin{aligned}
\alpha^{HH}&(\omega_{0})=\Delta v^{2}\sum_{i,j}^{N}\Big\lbrace \chi_{m}(\omega_{0})\mathbf{K}^{HH}(\mathbf{r}_{i},\mathbf{r}_{j},\omega_{0})\\[1pt]
&-\dfrac{ik_{0}}{2}\chi_{e}(\omega_{0})\mathbf{r}_{i}\wedge\mathbf{K}^{EH}(\mathbf{r}_{i},\mathbf{r}_{j},\omega_{0})\Big\rbrace e^{\im\mathbf{k}\cdot\mathbf{r}_{j}}\!.
\end{aligned}
\end{equation}
\end{subequations}
To be more precise, these are {\it pseudo-polarizabilities} since they depend on the direction of illumination due to the phase term $\exp(\im\mathbf{k}\cdot\mathbf{r}_{j})$.
Conventional polarizabilities depend only on the geometry and the material of the nanostructure.\cite{arangoPolarizabilityTensorRetrieval2013, bernalarangoUnderpinningHybridizationIntuition2014}
This phase term is the direct cause of the emergence of polarization vortices, which are responsible for the existence of magnetic multipole moments in dielectric nanostructures.\cite{kuznetsovMagneticLight2012, kuznetsovOpticallyResonantDielectric2016}
In order to be able to describe the magnetic polarization due to the mixed field susceptibility, we keep the phase term in the expression of the {\it pseudo polarizabilities}.
We note that this approximation is assuming plane wave illumination and requires that the wave vector of the incident field is known already during the calculation of \(\bm \alpha (\omega_0)\). 
However, we will show later, that a further approximation can be used to generalize these pseudo-polarizabilities to any oblique illumination without prior knowledge of the angle of incidence.
We note that it is possible to replace the phase term by an evanescent field, which however would lead to some loss of generality concerning the geometric orientation with respect to the incident field. 
Finally, fields like for instance a tightly focused Gaussian beam, can often be described as a series of plane waves, in which case the pseudo-polarizabilities can be applied without further modification of the formalism.

For the calculation of the polarizabilities we used our own python implementation ``pyGDM'' of the volume discretization procedure described above.\cite{wiechaPyGDMPythonToolkit2018}

\begin{figure*}[tp]
	\centering
	\includegraphics[width=\textwidth]{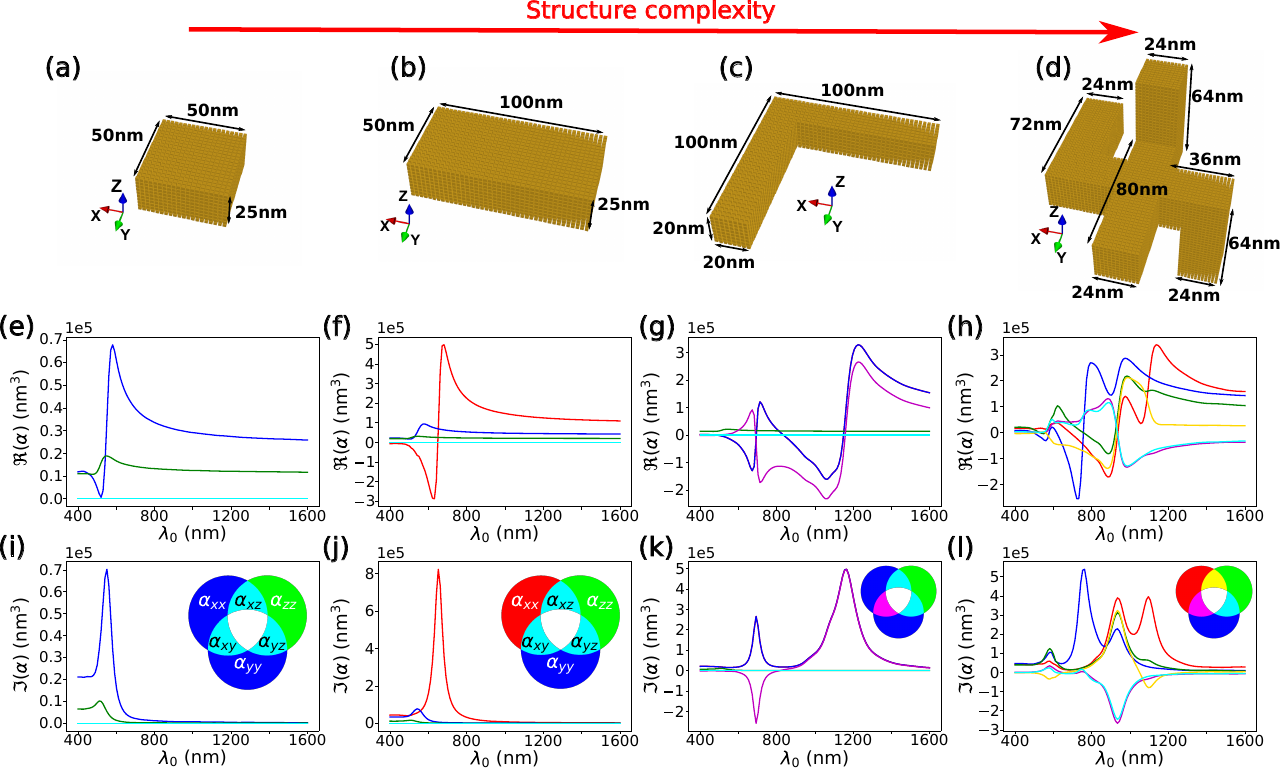}
	\caption{
	Spectral variation of the real (e)-(h) and imaginary part (i)-(l) of the terms of the dipolar polarizability matrix for various structures represented in (a)-(d). 
	Geometries consist in (a) a single isotropic pad of size $(50\text{nm}\times50\text{nm}\times25\text{nm})$, (b) an anisotropic pad of size $(50\text{nm}\times100\text{nm}\times25\text{nm})$, (c) an "L" shape structure included in the xOy plane, (d) a "3D" shape structure with ramifications in the three directions of space. 
	For the first three structures computations were performed with a discretization step d=2.5nm while we used a step d=2nm in the last case for a good convergence of the calculation.
	The color diagrams show the degenerate components of the polarizability tensor.
}
	\label{fig:complex_structures}
\end{figure*}

\section{Results}
\subsection{Electric-electric polarizability for structures of arbitrary shape}
In a first step, we compare the spectral variation of the imaginary part of the dipolar polarizability $\Im[\alpha^{EE}(\omega_{0})]$ at the example of an isolated spherical gold particle (radius \(r=5\,\)nm).
Fig.~\ref{fig:sphere_polarizability}a shows a comparison of the first diagonal term \(\alpha_{xx}\), calculated analytically (Eq.~(\ref{ALPHA1}), blue line) or numerically (using Eq.~(\ref{alpha_EE}), red line). For the sphere suspended in vacuum, the diagonal terms of \(\alpha^{EE}\) are identical, and off-diagonal terms vanish.
Our numerical discretization approach reproduces the well-known plasmon resonance for gold nano-particles around $\lambda_{0}=520$\,nm.\cite{amendolaSurfacePlasmonResonance2017}
The slight quantitative difference between the two representations is due to the inaccuracy of the analytical formula on non-atomic size scales.
If we add a silica substrate in the calculation (see inset in Fig.~\ref{fig:sphere_polarizability}b), the symmetry is reduced from spherical to a cylindrical. 
In consequence, the polarizability tensor is no longer diagonal and \(\alpha^{EE}_{xx}=\alpha^{EE}_{yy} \neq \alpha^{EE}_{zz}\), which is depicted in Fig.~\ref{fig:sphere_polarizability}b.
Here we use a Green's tensor based on the mirror charges technique to take into account the substrate, which is a quasistatic approximation. 
However, in our method a fully retarded Green's dyad can also be used to improve the accuracy for larger particles on higher index or plasmonic substrates.\cite{gay-balmazValidityDomainLimitation2000}
We show a comparison of quasistatic and fully retarded calculation of the polarizability in appendix~\ref{appendix:retardation}.
In appendix~\ref{appendix:mie_theoyr} we show furthermore a comparison of a larger gold nanosphere to Mie theory.

The volume discretization allows us to treat nanostructures of arbitrary shape.\cite{girardFieldsNanostructures2005, wiechaPyGDMPythonToolkit2018}
Therefore, in a next step we study the evolution of the different terms of the {\it electric-electric pseudo-polarizability} tensor $\alpha^{EE}(\omega_{0})$, while gradually increasing the structure complexity, as illustrated in Fig.~\ref{fig:complex_structures}(a-d).
Note that the polarizability tensor is symmetric (see Eqs.~\eqref{alpha_EE} and~\eqref{MIXED-S}), so in Fig.~\ref{fig:complex_structures}(e-l) we plot only the upper triangular elements.
First, we calculate the spectral variation of the polarizability matrix of a gold pad of size $(50\text{nm}\times50\text{nm}\times25\text{nm})$, discretized with cubic cells of side length \(d=2.5\)\,nm (cf. Fig. \ref{fig:complex_structures}a).
The real and imaginary part of each tensor component are shown in Fig.~\ref{fig:complex_structures}e, respectively~\ref{fig:complex_structures}i.
Due to the symmetry of the structure the off diagonal terms of \(\alpha^{EE}\) are zero (cyan lines).
Moreover, we observe that \(\alpha^{EE}_{xx}= \alpha^{EE}_{yy}\) (blue lines) which is a result of the rectangular footprint of the structure. 
Because the height is only half of the structure's width, \(\alpha^{EE}_{zz}\) is significantly smaller (green line).
Despite the small dimensions of the pad, localized plasmon resonances arise slightly red-shifted at around \(550 - 600\,\)nm.
Now if we increase the size of the pad along \(Ox\) by a factor of two, the  \(\alpha^{EE}_{xx}\) and \(\alpha^{EE}_{yy}\) terms are not equal anymore, due to the aspect ratio of the elongated pad.
In this case, the resonance for excitation along the long edge is even more red-shifted to around \(650\)\,nm, which reflects the effective wavelength scaling of the localized plasmon resonance.\cite{novotnyEffectiveWavelengthScaling2007}
Next, we calculate the polarizability tensor for a symmetric L-shaped gold structure (illustrated in Fig.~\ref{fig:complex_structures}c). In this structure, coupling between the horizontal and the vertical arm leads to a non-zero off-diagonal term \(\alpha^{EE}_{xy}\), as can be seen in Fig.~\ref{fig:complex_structures}g and~\ref{fig:complex_structures}k (magenta lines).
Due to this off-diagonal term, two additional resonances emerge around 690\,nm and 1170\,nm at which polarization conversion between the \(Y\)-arm and the \(X\)-arm of the antenna occurs.\cite{katsGiantBirefringenceOptical2012, wiechaPolarizationConversionPlasmonic2017}
The two peaks at \(690\,\)nm and \(1170\,\)nm correspond to the anti-bonding, respectively bonding modes between the two arms.\cite{panaroDarkBrightMode2014, blackOptimalPolarizationConversion2014}
We note, that the opposite phase of the bonding and the anti-bonding mode is correctly reflected also in the spectrum of the polarizability off-diagonal element.
Polarization conversion is only occurring between \(X\) and \(Y\), hence the other off-diagonal elements remain zero (cyan lines).
Moreover, both arms are of the same length which leads to \(\alpha^{EE}_{xx}= \alpha^{EE}_{yy}\) (blue lines).
Finally, we construct a three-dimensional structure which introduces interactions between each Cartesian direction, as depicted in figure~\ref{fig:complex_structures}d.
In this case, each matrix element shows a unique spectral behavior, representing the complex interaction mechanisms between the antenna arms in different directions (Fig. \ref{fig:complex_structures}h and \ref{fig:complex_structures}l).

While the effective polarizability approximation is mainly interesting for the description of far-field characteristics where the dipolar response usually dominates, it can also be used to a certain extent to calculate the electromagnetic field in the vicinity of a nanostructure. 
However, as shown in appendix~\ref{appendix:near-field}, the accuracy in the near-field decreases dramatically if the field is to be evaluated too close to the nanostructure or when the local optical response cannot be described by a single dipolar point-source.
In case of static polarizabilities (neglecting the phase term in Eqs.~\eqref{eq:different_alphas}) it is furthermore possible to re-introduce optical interactions between several polarizabilities via a coupling scheme as used in the Green's Dyadic Method (GDM).\cite{martinGeneralizedFieldPropagator1995, wiechaPyGDMPythonToolkit2018} 
We demonstrate this in appendix~\ref{appendix:nearfield_coupling}, where we also discuss the limitations of the coupled effective polarizability model in terms of minimum inter-particle distances and near-field accuracy.

We note at this point, that the approach is also capable to deal with nano-cavities carved into a bulk medium, by using a non-unitary permittivity for the environment and \(\chi_e=\chi_m=0\) in the hollow source region.

\begin{figure}[t]
	\centering\includegraphics[width=\columnwidth]{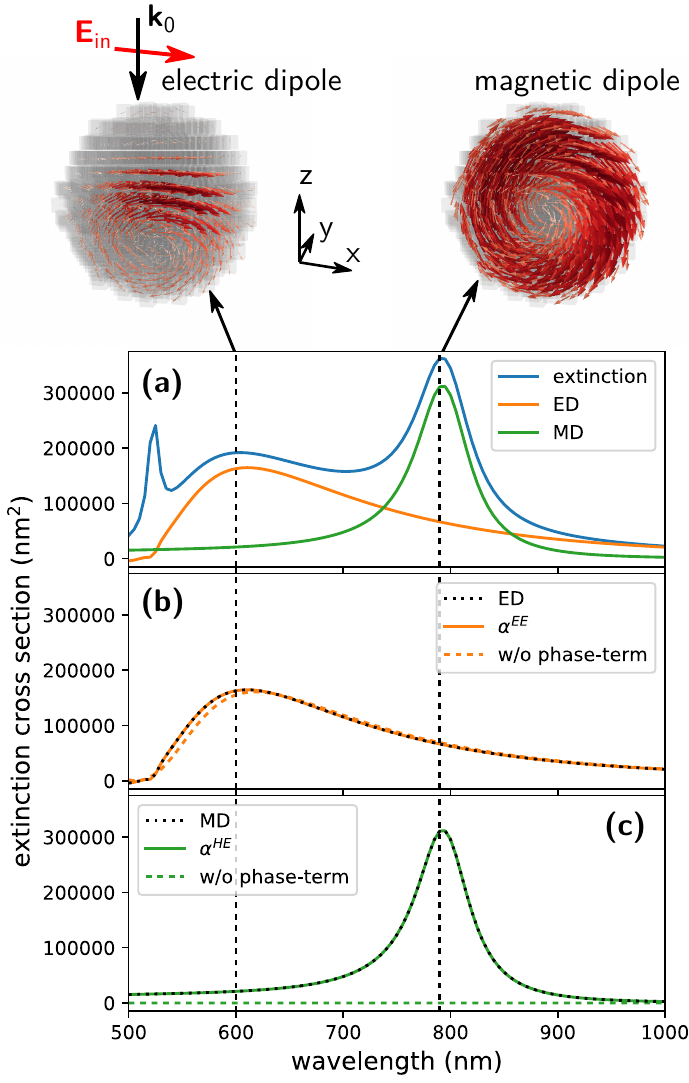}
	\caption{
		Dielectric nanosphere (\(n=4\)) of radius \(r=100\,\)nm in vacuum, illuminated by a plane wave of linear polarization.
		(a) Extinction spectrum (blue line) and electric (ED) as well as magnetic dipole (MD) contributions to the extinction (orange, respectively green line), calculated as described in Ref.~\onlinecite{evlyukhinMultipoleLightScattering2011}.
		(b) Electric dipole extinction via the polarizability tensor \(\alpha^{EE}\) with (solid orange line) and without the phase term (dashed orange line).
		The ED extinction from (a) is shown as dashed black line for comparison.
		(c) Magnetic dipole extinction via the polarizability tensor \(\alpha^{HE}\) with (solid green line) and without the phase term (dashed green line).
		The MD extinction from (a) is shown as dashed black line for comparison.
		At the top, the internal electric field distribution (real parts) is shown at the ED (left) and MD resonance (right).
	}
	\label{fig:alpha_EH}
\end{figure}

\begin{figure*}[t]
	\centering\includegraphics[width=\textwidth]{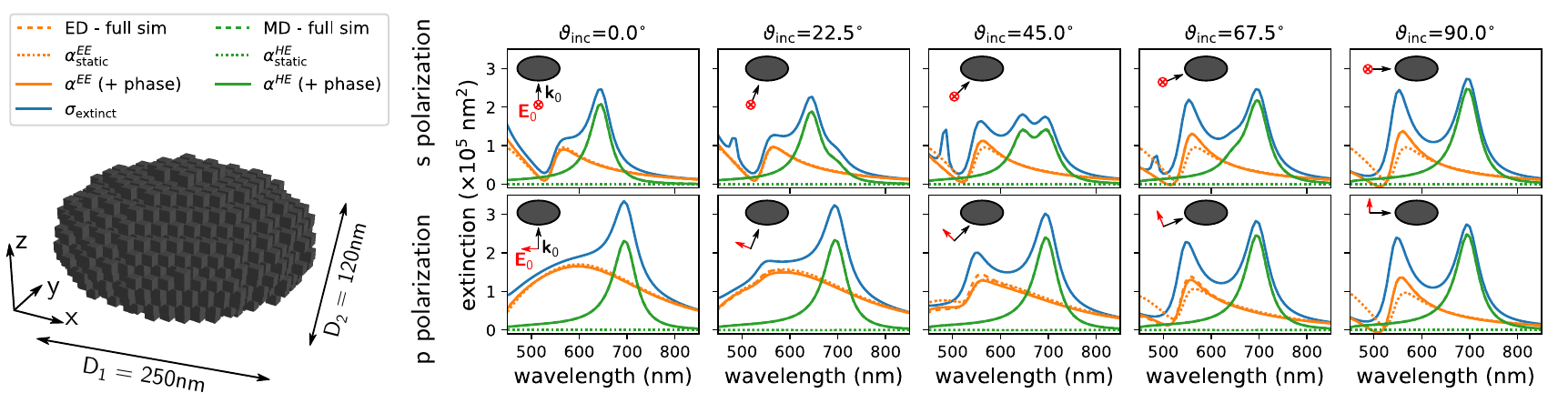}
	\caption{
		Dielectric nano-spheroid (\(n=4\)) of long axis diameter \(D_1=250\,\)nm (along \(Ox\)) and short axis diameter \(D_2=120\,\)nm (along \(Oy\) and \(Oz\)) placed in vacuum, illuminated by a plane wave of linear polarization (top: \(s\)-polarization, bottom: \(p\)-polarization) for varying incident angles \(\vartheta_{\text{inc}}\).
		The total extinction cross section (blue solid line) is compared to the extinction induced by the effective electric and magnetic dipole moments, obtained through full-field simulations (dashed orange and green lines), to the pseudo-polarizability (solid orange and green lines) as well as to the ``static'' polarizabilities (dotted orange and green lines).
		The pseudo-polarizability superposition approximation is used in the three cases of oblique incidence (\(22.5^{\circ}\), \(45^{\circ}\) and \(67.5^{\circ}\)).
	}
	\label{fig:interpolation_k}
\end{figure*}

\subsection{Magnetic-electric polarizability of a dielectric sphere}

We now want to assess the role of the magnetic terms in the super polarizability.
Since in nature no material with a significant direct magnetic optical response is known, we will assume \(\chi_m = 0\), hence the magnetic field of light cannot directly interact with the nanostructure.
In consequence the polarizability tensors Eqs.~(\ref{eq:different_alphas}) drastically simplify.
The mixed terms involving \(\mathbf{K}^{EH}\) and \(\mathbf{K}^{HE}\) all disappear, since they include the product \(\chi_e \chi_m\) (see also Eq.~\eqref{MIXED-S}). 
In fact only the two terms that depend on \(\mathbf{K}^{EE}\) remain.
Hence, for media with \(\chi_m=0\), the electric polarization is fully described by \(\alpha^{EE}\) and the magnetic polarization is entirely governed by \(\alpha^{HE}\).

As an example we show in figure~\ref{fig:alpha_EH}a the extinction cross section of a dielectric nano-sphere (\(n=4\)) of radius \(r=100\,\)nm in vacuum, calculated from the discretized electric polarization density (blue line).\cite{girardFieldsNanostructures2005}
We show additionally the decomposition of the extinction into an effective electric and magnetic dipole moment at the sphere's center of mass (orange, respectively green lines).\cite{evlyukhinMultipoleLightScattering2011, terekhovMultipolarResponseNonspherical2017}
The dielectric sphere has an electric dipole (ED) resonance at \(600\,\)nm and a magnetic dipole (MD) resonance at \(790\,\)nm, which are indicated by black vertical dashed lines. 
The real part of the electric field inside the nano-sphere at these resonances is qualitatively shown in 3D vector plots above figure~\ref{fig:alpha_EH}a.
In figure~\ref{fig:alpha_EH}b and~\ref{fig:alpha_EH}c we show the extinction cross section obtained from the effective polarizabilities \(\alpha^{EE}\), respectively \(\alpha^{HE}\). 
We compare the ``static'' effective polarizabilities without phase term (dashed colored lines) and the above introduced pseudo-polarizabilities including the phase term \(\exp(i\,\mathbf{k}\cdot \mathbf{r}_j)\) (solid colored lines, see Eqs.~\eqref{eq:different_alphas}). 
The dotted black lines show the ED and MD response from the full internal fields.
While the ED resonance in figure~\ref{fig:alpha_EH}b is very well reproduced by both, the static and the phase-sensitive electric-electric pseudo-polarizability, the MD resonance cannot be reproduced if the phase term in Eq.~\eqref{eq:alpha_HE} is omitted (dashed green line in figure~\ref{fig:alpha_EH}c). 
Only if the phase term is taken into account, the extinction calculated from the pseudo-polarizability matches the magnetic dipole resonance in the dielectric sphere (solid green line in figure~\ref{fig:alpha_EH}c).
This is because the magnetic dipole is induced by the vortex formed by the electric displacement current (see illustration of the MD above Fig.~\ref{fig:alpha_EH}a, right), which is a direct consequence of the phase difference of the incident field across the relatively large nano-sphere.

To test our model for energy conservation, we show in appendix~\ref{appendix:ext_scat} a comparison of extinction and scattering cross sections at the example of a \(r=100\,\)nm and \(n=4\) lossless, dielectric nanosphere.


\subsection{Approximation of \(\alpha^{HE}\) for arbitrary angles of incidence}

In contrast to ``classical'' static polarizabilities,\cite{sersicMagnetoelectricPointScattering2011, arangoPolarizabilityTensorRetrieval2013} the here introduced \textit{pseudo polarizabilities} depend on the illumination wave-vector \(\mathbf{k}_0\) as a result of the above discussed phase term.
In consequence, to solve the general problem, the pseudo polarizability needs to be separately calculated for every incident field which limits the usefulness of the approximation.
However, we can approximate arbitrary incident angles through a first order expansion of the phase term. 
While we keep the phase-term in the definition of the polarizabilities, we assume that the first order term of its Taylor expansion is sufficient to describe the magnetic dipolar response.
Thus, while allowing retardation effects to a certain extent, we still stick with the assumption that the wavelength is large with respect to the nanostructure (i.e. \(\lambda_0 \gg |\mathbf{r}|\)). 
Since the optical interaction is still modelled as a point-response, the wave vector of the illumination is assumed to be constant across the nanostructure. 
Furthermore, the approximation requires that the location of the effective dipole is independent of the wave vector.
We assume here that the effective electric and magnetic dipole moments \(\mathcal{P}(\omega_{0})\), respectively \(\mathcal{M}(\omega_{0})\) lie at the particle's center of mass \(\mathbf{r}_c\) for any angle of incidence and polarization of the illumination.
Without loss of generality we now consider an incident wave vector in the \(XZ\) plane, were we get:
\begin{subequations}\label{eq:polarizations_sum_wavevector}
\begin{equation}
\begin{split}
\mathcal{P}\ \approx\ 
	\Bigg( \left(\dfrac{k_x}{|\mathbf{k_0}|}\right)^2 \alpha^{EE}_{k_x} +\ 
	\left(\dfrac{k_z}{|\mathbf{k_0}|}\right)^2 \alpha^{EE}_{k_z} \Bigg) 
	\cdot \mathbf{E}_0
\end{split}
\end{equation}
\text{and}
\begin{equation}
\begin{split}
\mathcal{M}\ \approx\ 
\bigg( \dfrac{k_x}{|\mathbf{k_0}|} \alpha^{HE}_{k_x} +\ 
\dfrac{k_z}{|\mathbf{k_0}|} \alpha^{HE}_{k_z} \bigg) 
\cdot \mathbf{E}_0
	\, .
\end{split}
\end{equation}
\end{subequations}
For a derivation of these approximations based on a first order expansion of the phase term \(\exp(\text{i}\mathbf{k}\cdot\mathbf{r}_j)\) in Eqs.~(\ref{eq:different_alphas}), see appendices~\ref{appendix:alpha_HE} and~\ref{appendix:alpha_EE}.
The dependence on \(\omega_0\) and \(\mathbf{r}_c\) has been omitted for the sake of readability. \(k_i\) is the wave vector component and \(\alpha^{XX}_{k_i}\) the pseudo polarizability for the Cartesian direction \(i\in \{x, z\}\). 
Both are evaluated at the position \(\mathbf{r}_c\) of the effective dipole (here the center of mass).
Using this superposition scheme, the response of the nanostructure to any oblique plane wave illumination is described by three ``pseudo super-polarizability tensors'' \(\bm{\alpha}_{k_i}\) (one for every Cartesian coordinate axis $i$).
Once the approximations Eqs.~\eqref{eq:polarizations_sum_wavevector} for the effective dipole moments are calculated, the extinction cross sections due to the induced electric and magnetic polarizations can be calculated as\cite{evlyukhinMultipoleLightScattering2011}

\begin{subequations}\label{eq:extinction_arbitrary_incidence}
\begin{equation}
I_{\text{ext},\mathcal{P}}(\omega_0) =  \frac{8 \pi^{2}}{n_1 \lambda_{0}} \ \Im \Big( \mathbf{E}_0(\mathbf{r}_c, \omega_0)^* \cdot \mathcal{P}(\omega_0) \Big)
\end{equation}
\begin{equation}
I_{\text{ext},\mathcal{M}}(\omega_0) =  \frac{8 \pi^{2}}{n_1 \lambda_{0}} \ \Im \Big( \mathbf{H}_0(\mathbf{r}_c, \omega_0)^* \cdot \mathcal{M}(\omega_0) \Big)
\, ,
\end{equation}
\end{subequations}
where the superscript asterisk (\(^*\)) indicates complex conjugation.

In figure~\ref{fig:interpolation_k} we show spectra of the extinction cross section of a dielectric spheroid (refractive index \(n=4\)) in vacuum, with a diameter of \(D_1=250\,\)nm along the \(OX\)-oriented long axis and two identical short axes with diameters (\(D_2=120\,\)nm), as illustrated at the left of figure~\ref{fig:interpolation_k}.
The extinction is shown for different incident angles for \(s\)-polarization (top row) and \(p\)-polarization (bottom row).
Clearly, the pseudo polarizability superposition approximation (solid lines; \(\alpha^{EE}\): orange,  \(\alpha^{HE}\): green) yields excellent agreement with the ED and MD decomposition of the extinction from full-field simulations (dashed orange and green lines, respectively).
Once again, the static polarizability approximation breaks down in case of the magnetic dipole resonance \(\alpha^{HE}_{\text{static}}\) (green dotted lines). 
In case of the electric dipole response, the static polarizability \(\alpha^{EE}_{\text{static}}\) gives a reasonable approximation. 
However, if the incidence direction is along the long axis of the ellipsoid, phase effects start to play a non-negligible role, and significant deviations occur in the static polarizability approximation.

Despite the size of the nanostructure seemingly outside the range of validity of the first order expansion of the exponential in Eqs.~\eqref{eq:different_alphas}, we observe an excellent agreement with the full simulation.
We recall that the pseudo-polarizability approximation is assuming an effective, punctual optical response at the center of mass of the nanostructure. 
At this specific position we then superpose the effective dipoles for different angles of incidence.
As long as the effective position of the electric dipole moment stays at the center of mass and the field vortex spins around the center of mass position, the region where we apply our superposition scheme for different angles of incidence is confined to a volume where \(kD\ll 1\).
We observe that in cases of more irregularly shaped nanostructures than the above ellipse, the effective positions of the electric and magnetic dipole moment are not necessarily at or even close to the center of mass, and can furthermore vary significantly with the angle of incidence.
We therefore show in the appendix~\ref{appendix:large_dielectric} simulations of two less symmetric dielectric nanostructures under oblique incidence, which still show very good agreement to full field simulations, but in which inaccuracies in the superposition approximation start to occur.
In various tests we observed that before the superposition approximation would lead to significant errors, the dipolar approximation breaks down as a result of the occurrence of higher order modes.

\section{Conclusion and perspectives}
In conclusion we introduced a mathematical scheme for a generalized description of light-matter interaction in nanostructures through both, optical electric and magnetic fields. 
We showed how the optical response of nanostructures can be approximated through a universal ``super polarizability'' tensor, which combines the optical response through electric and magnetic dipole moments.
Using a volume discretization, the super polarizability can be numerically calculated for nano-structures of arbitrary shape and material.
We demonstrated that our pseudo polarizability, which includes phase effects, is capable to faithfully describe also magnetic dipole resonances in dielectric nanostructures of important size, where a conventional, static point-response model is breaking down.
In contrast to similar, computationally more complex multi-dipole methods,\cite{bertrandGlobalPolarizabilityMatrix2019} our approach of effective electric and magnetic polarizabilities is capable to capture the optical response of complex nanostructures in a single quantity, which strongly facilitates the further evaluation of the optical behavior, for instance under changing illumination conditions.
We foresee that our framework can be used to calculate large assemblies of different and / or randomly positioned nanostructures. 
Our work will therefore be very useful in the simulation of highly heterogeneous, non-periodic assemblies of plasmonic nanostructures and will be helpful also for the description of weakly coupled assemblies of dielectric nanostructures in the Born approximation. 
It might pave the way to the development of design methods for complex, non-periodic, hybrid metasurfaces.

\bigskip
\noindent
{\bf Acknowledgments}:
The authors thank G. Colas des Francs and A. Mlayah for fruitful discussions on the physics of small plasmonic particles.
PRW acknowledges support by the German Research Foundation (DFG) through a research fellowship (WI 5261/1-1).
This work was supported by the computing center CALMIP in Toulouse.
All data supporting this study are openly available from the University of Southampton repository (DOI: 10.5258/SOTON/D1177).
\par
\bigskip
\noindent

\appendix
\section*{Appendix}
\renewcommand{\thesubsection}{\Alph{subsection}}

\subsection{Interpolation of magnetic-electric polarizabilities}\label{appendix:alpha_HE}
For a non-magnetic nanostructure the electric-magnetic polarizability writes (see also main paper):
\begin{equation}
\begin{aligned}
\alpha^{HE} (\omega_{0}) = &
\sum_{i,j}^{N} 
-\dfrac{ik_{0}}{2}\chi_{e} \Delta v^{2}
\Big\lbrace
\mathbf{r}_{i}\wedge\mathbf{K}^{EE}(\mathbf{r}_{i},\mathbf{r}_{j})
\Big\rbrace e^{\im \mathbf{k}\cdot\mathbf{r}_{j}} \\
= &
\sum_{i,j}^{N} A_{i,j}^{HE} e^{\im\mathbf{k}\cdot\mathbf{r}_{j}} \, ,
\end{aligned}
\end{equation}
where we neglected the dependence on \(\omega_0\) for the sake of readability.
Due to the phase term \(\exp(\im \mathbf{k}\cdot \mathbf{r})\), the polarizability \(\alpha^{HE}\) is dependent on the incident angle and writes for a wave vector \(\mathbf{k}\) of arbitrary angle \(\vartheta\):
\begin{equation}
\alpha_{\vartheta}^{HE}=\sum_{i,j}^{N} A_{i,j}^{HE} e^{\im\mathbf{k}\cdot\mathbf{r}_{j}} =\sum_{i,j}^{N} A_{i,j}^{HE} e^{\im(k_x x_{j} + k_y y_{j} + k_z z_{j})}\, .
\end{equation}
Now we consider three $\alpha^{HE}_{r_i}$ corresponding to plane wave incidence along each of the three Cartesian directions: 
\begin{equation}
\begin{aligned}
\alpha_{x}^{HE} &= \sum_{i,j}^{N} A_{i,j}^{HE} e^{\im k x_{j}}\\
\alpha_{y}^{HE} &= \sum_{i,j}^{N} A_{i,j}^{HE} e^{\im k y_{j}}\\
\alpha_{z}^{HE} &= \sum_{i,j}^{N} A_{i,j}^{HE} e^{\im k z_{j}};
\end{aligned}
\end{equation}
where $k = \dfrac{2\pi n}{\lambda_0}$ and $n$ is the medium index.
\par\leavevmode\par
We now develop the sum of the polarizabilities for plane wave incidence along the Cartesian coordinate axis.
We define also three parameters allowing to describe an arbitrary illumination direction :
\begin{equation}
\label{eq:beta_definition}
\beta_x = \dfrac{k_x}{ k}\, ,
\quad
\beta_y = \dfrac{k_y}{ k}\, ,
\quad
\beta_z = \dfrac{k_z}{ k}\, ,
\end{equation} 
In addition, we assumed that \(k_x^2+k_y^2+k_z^2 = k^2\). We can now write
\begin{multline}
\beta_x\alpha_x^{HE} + \beta_y\alpha_y^{HE} + \beta_z\alpha_z^{HE}
= \\
\sum_{i,j}^{N} A_{i,j}^{HE}
\Big[\beta_x e^{\im k x_{j}} + \beta_y e^{\im k y_{j}} + \beta_z e^{\im k z_{j}}\Big]\, .
\end{multline}
Assuming that \(\lambda_0 \gg |\mathbf{r}| \), we can approximate the exponentials by their first order Taylor series:
\begin{equation}
\begin{aligned}
& \beta_x\alpha_x^{HE} + \beta_y\alpha_y^{HE} + \beta_z\alpha_z^{HE}
\approx \\
& \sum_{i,j}^{N} A_{i,j}^{HE}\Big[ \beta_x(1 + i k x_{j}) + \beta_y(1 + i k y_{j}) + \beta_z(1 + i k z_{j})\Big] = \\
& \sum_{i,j}^{N} A_{i,j}^{HE}\Big[\beta_x + \beta_y + \beta_z + \im(\beta_x  k x_{j} + \beta_y  k y_{j} + \beta_z  k z_{j})\Big]\, .
\end{aligned}
\end{equation}
By adding ``$1-1$'', we can write
\begin{equation}\label{eq:development_aHE_with_constants}
\begin{aligned}
&\beta_x\alpha_x^{HE} + \beta_y\alpha_y^{HE} + \beta_z\alpha_z^{HE}
\approx \\
&\sum_{i,j}^{N} A_{i,j}^{HE}[\beta_x + \beta_y + \beta_z - 1 + \underbrace{1 + i(k_x x_{j} + k_y y_{j} + k_z z_{j})}_{\approx\ e^{\im(k_x x_{j} + k_y y_{j} + k_z z_{j})}}] \approx  \\
& \sum_{i,j}^{N} A_{i,j}^{HE}[\beta_x + \beta_y + \beta_z - 1 + e^{\im(k_x x_{j} + k_y y_{j} + k_z z_{j})}]\, .
\end{aligned}
\end{equation}
The constant terms in Eq.~\eqref{eq:development_aHE_with_constants} are proportional to the static magnetic-electric polarizability, which, as we have shown in the main paper, is negligible compared to usual dipolar polarizabilities, since the vortices that generate the magnetic dipolar response in non-magnetic nanostructures cannot be described without the phase term \(\exp(\im \mathbf{k}\cdot \mathbf{r})\):
\begin{equation} \label{eq:magnetic_0_pola_constant}
\sum_{i,j}^{N} \Big( \text{const.} \times \, A_{i,j}^{HE} \Big) \approx 0\, .
\end{equation}
Hence we find:
\begin{equation}
\alpha_{\vartheta}^{HE}
\approx 
\beta_x\alpha_x^{HE} + \beta_y\alpha_y^{HE} + \beta_z\alpha_z^{HE}
\end{equation}

\subsection{Interpolation of electric-electric polarizabilities}\label{appendix:alpha_EE}

The electric-electric polarizability writes (see also main paper):
\begin{equation}
\begin{aligned}
\alpha^{EE}(\omega_{0}) = &
\Delta v^{2}\chi_{e}(\omega_{0})
\sum_{i,j}^{N}
\mathbf{K}^{EE}(\mathbf{r}_{i},\mathbf{r}_{j},\omega_{0})
e^{i\mathbf{k}\cdot\mathbf{r}_{j}} \\
= &
\sum_{i,j}^{N} A_{i,j}^{EE} e^{\im\mathbf{k}\cdot\mathbf{r}_{j}} \, ,
\end{aligned}
\end{equation}
where we neglected the dependence on \(\omega_0\) for the sake of readability.

Due to the phase term \(\exp(\im \mathbf{k}\cdot \mathbf{r})\), the polarizability \(\alpha^{EE}\) is dependent on the incident angle and writes for a wave vector \(\mathbf{k}\) of arbitrary angle \(\vartheta\):
\begin{equation}\label{eq:alpha_teta_EE_general_k}
\alpha_{\vartheta}^{EE}=\sum_{i,j}^{N} A_{i,j}^{EE} e^{\im\mathbf{k}\cdot\mathbf{r}_{j}} =\sum_{i,j}^{N} A_{i,j}^{EE} e^{\im(k_x x_{j} + k_y y_{j} + k_z z_{j})}\, .
\end{equation}
Now we consider three $\alpha^{EE}_{r_i}$ corresponding to plane wave incidence along each of the three Cartesian directions: 
\begin{equation}
\begin{aligned}
\alpha_{x}^{EE} &= \sum_{i,j}^{N} A_{i,j}^{EE} e^{\im k x_{j}}\\
\alpha_{y}^{EE} &= \sum_{i,j}^{N} A_{i,j}^{EE} e^{\im k y_{j}}\\
\alpha_{z}^{EE} &= \sum_{i,j}^{N} A_{i,j}^{EE} e^{\im k z_{j}}
\end{aligned}
\end{equation}
We use the same definition of $\beta_x$, $\beta_y$ and $\beta_z$ as in Eq.~\eqref{eq:beta_definition}.
We can now write:
\begin{multline}
\beta_x^2\alpha_x^{EE} + \beta_y^2\alpha_y^{EE} + \beta_z^2\alpha_z^{EE}
= \\
\sum_{i,j}^{N} A_{i,j}^{EE}
\Big[\beta_x^2 e^{\im k x_{j}} + \beta_y^2 e^{\im k y_{j}} + \beta_z^2 e^{\im k z_{j}}\Big]\, .
\end{multline}
Assuming that \(\lambda_0 \gg |\mathbf{r}| \), we can approximate the exponentials by their first order Taylor series:
\begin{equation}\label{eq:apprixmate_sum_alphaEE}
\begin{aligned}
& \beta_x^2\alpha_x^{EE} + \beta_y^2\alpha_y^{EE} + \beta_z^2\alpha_z^{EE} \approx \\
& \sum_{i,j}^{N} A_{i,j}^{EE}\Big[ \beta_x^2(1 + i k x_{j}) + \beta_y^2(1 + i k y_{j}) + \beta_z^2(1 + i k z_{j})\Big] = \\
& \sum_{i,j}^{N} A_{i,j}^{EE}\Big[\underbrace{\beta_x^2 + \beta_y^2 + \beta_z^2}_{=1} + \im(\beta_x^2  k x_{j} + \beta_y^2  k y_{j} + \beta_z^2  k z_{j})\Big] = \\
&  \sum_{i,j}^{N} A_{i,j}^{EE}\Big[1 + \im(\beta_x^2  k x_{j} + \beta_y^2  k y_{j} + \beta_z^2  k z_{j})\Big]\, .
\end{aligned}
\end{equation}
Now we subtract the first order Taylor expansion of equation~\eqref{eq:alpha_teta_EE_general_k} 
\begin{equation}
\alpha_{\vartheta}^{EE}
\approx \sum_{i,j}^{N} A_{i,j}^{EE}\Big[1 + \im(\beta_x  k x_{j} + \beta_y  k y_{j}+ \beta_z  k z_{j})\Big]
\end{equation}
from both sides of equation~\eqref{eq:apprixmate_sum_alphaEE}, which yields:
\begin{equation} \label{eq:final_difference_expression_aEE}
\begin{aligned}
& \beta_x^2\alpha_x^{EE} + \beta_y^2\alpha_y^{EE} + \beta_z^2\alpha_z^{EE} -\alpha_{\vartheta}^{EE}
\approx \\
& \sum_{i,j}^{N} \im A_{i,j}^{EE} \Big(\beta_x(\beta_x - 1)  k x_{j} + \beta_y(\beta_y -1)  k y_{j} + \beta_z(\beta_z-1)  k z_{j}\Big) = \\
& \qquad \quad \im \beta_x(\beta_x - 1)  k \sum_{i,j}^{N} A_{i,j}^{EE} x_{j} \\
& \qquad + \quad \im \beta_y(\beta_y - 1)  k \sum_{i,j}^{N} A_{i,j}^{EE} y_{j} \\
& \qquad + \quad \im \beta_z(\beta_z - 1)  k \sum_{i,j}^{N} A_{i,j}^{EE} z_{j}
\end{aligned}
\end{equation}
To demonstrate that the expression on the right hand side in Eq.~\eqref{eq:final_difference_expression_aEE} is negligible, we use Eq.~\eqref{eq:magnetic_0_pola_constant}, which states that
\begin{multline}
\sum_{i,j}^{N} \Big(A_{i,j}^{HE} \Big)
= \\
- \Delta v^{2}\dfrac{ik_{0}}{2}\chi_{e}(\omega_{0})\sum_{i,j}^{N}\Big\lbrace\mathbf{r}_{i}\wedge\mathbf{K}^{EE}(\mathbf{r}_{i},\mathbf{r}_{j},\omega_{0})\Big\rbrace \approx 0\, .
\end{multline}
Properly speaking, with ``approximately zero'' we mean that the term is negligible within the small particle approximation.
Using now the symmetry of $\mathbf{K}^{EE}(\mathbf{r}_{i},\mathbf{r}_{j},\omega_{0})$ and the antisymmetry of the cross product ``$\mathbf{r}_{i}\, \wedge$'', we can anti-commute those two terms:
\begin{equation}
\begin{aligned}
\sum_{i,j}^{N} \Big(A_{i,j}^{HE} \Big)
&= \Delta v^{2}\dfrac{ik_{0}}{2}\chi_{e}(\omega_{0})\sum_{i,j}^{N}\mathbf{K}^{EE}(\mathbf{r}_{i},\mathbf{r}_{j},\omega_{0})\cdot\Big\lbrace\mathbf{r}_{i}\, \wedge \Big\rbrace\\
&= \dfrac{ik_{0}}{2}\sum_{i,j}^{N} A_{i,j}^{EE}\cdot\Big\lbrace\mathbf{r}_{i}\, \wedge \Big\rbrace \approx 0 
\end{aligned}
\end{equation}
where
\begin{equation}
\Big\lbrace\mathbf{r}_{i}\, \wedge\Big\rbrace = \begin{pmatrix}
0  & -z_i &  y_i \\
z_i &   0  & -x_i \\
-y_i &  x_i &   0  \\
\end{pmatrix} \, .
\end{equation}
Using the symmetry of \(\mathbf{K}^{EE}\) $\left(A_{i,j}^{EE} =  A_{j,i}^{EE}\right)$, we get
\begin{equation} \label{eq:decomposition_approx_terms_Aij_xj_0}
\begin{aligned}
\dfrac{ik_{0}}{2}\sum_{i,j}^{N} A_{i,j}^{EE} x_{j} & \approx 0\\
\dfrac{ik_{0}}{2}\sum_{i,j}^{N }A_{i,j}^{EE} y_{j} & \approx 0\\
\dfrac{ik_{0}}{2}\sum_{i,j}^{N} A_{i,j}^{EE} z_{j} & \approx 0
\end{aligned}
\end{equation}
Comparing Eq.~\eqref{eq:final_difference_expression_aEE} and Eq.~\eqref{eq:decomposition_approx_terms_Aij_xj_0} we find
\begin{equation}
(\beta_x^2\alpha_x^{EE} + \beta_y^2\alpha_y^{EE} + \beta_z^2\alpha_z^{EE}) - \alpha_{\theta}^{EE} 
\approx 0 \, ,
\end{equation}
hence
\begin{equation}
\alpha_{\vartheta}^{EE}
\approx 
\beta_x^2\alpha_x^{EE} + \beta_y^2\alpha_y^{EE} + \beta_z^2\alpha_z^{EE}\, .
\end{equation}

\begin{figure}[t!]
	\centering\includegraphics[width=0.85\columnwidth]{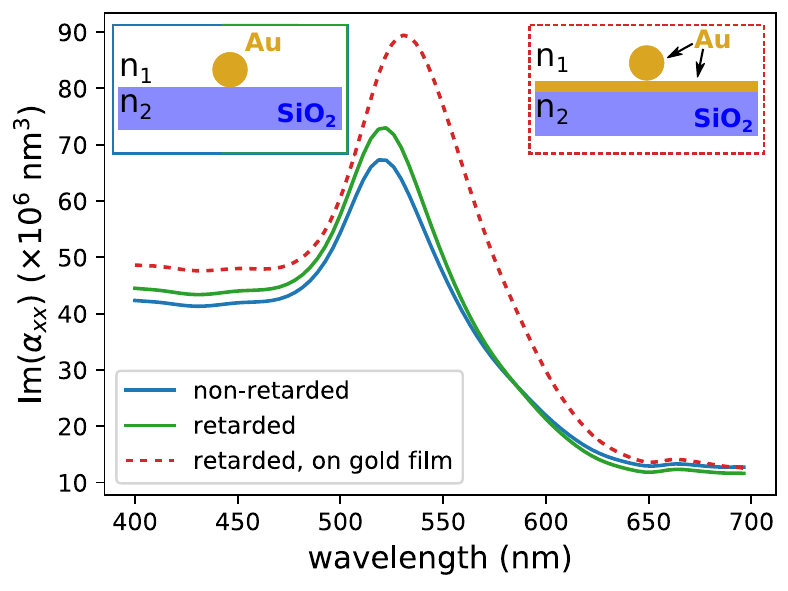}
	\caption{
		Polarizability of a gold sphere (radius \(r=50\,\)nm) calculated using a non-retarded (blue line) and a retarded (green and red lines) Green's tensor for the description of the substrate.
		The blue and green curves correspond to the \(\alpha_{xx}\) component of the electric-electric polarizability of the nano-sphere lying on a dielectric substrate (\(n_2=1.45\), illustrated in the top left inset). 
		In the case of the red dashed curve an additional \(50\,\)nm thick gold layer is inserted between silica substrate and gold sphere (see sketch in top right inset).
		In both cases, the top medium is air (\(n_1=1\)).
	}
	\label{fig:retard_non_retard}
\end{figure}
\begin{figure}[t!]
	\centering\includegraphics[width=\columnwidth]{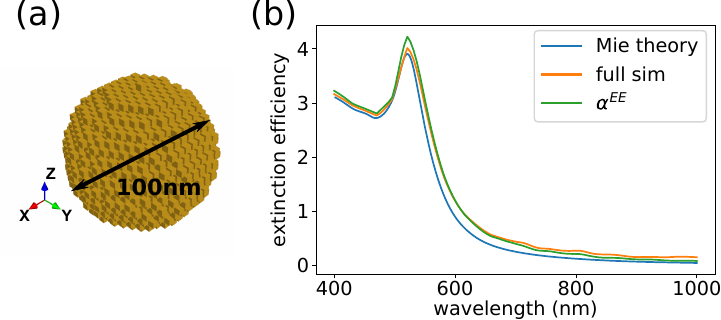}
	\caption{
		Extinction efficiency spectra computed for a gold sphere of diameter \(D=100\,\)nm using Mie theory (blue curve), the full GDM simulation (orange curve) and the effective polarizability approximation (green curve). 
		The sphere is placed in vacuum and illuminated by a linear polarized plane wave.
	}
	\label{fig:mie_comparison_goldsphere}
\end{figure}

\subsection{Particles on a substrate: quasistatic approximation vs. retardation}\label{appendix:retardation}

Using an appropriate Green's tensor, our approach permits to include a substrate for the extraction of the polarizability. 
The dipole moment of the polarizability, excited with an arbitrary illumination, then includes implicitly the optical interaction with the substrate.
For the calculation shown in the main text figure~\ref{fig:sphere_polarizability}b we used a Green's dyad based on a quasistatic mirror-charge approximation to describe the substrate.\cite{gay-balmazValidityDomainLimitation2000} 
To assess whether this is an appropriate approximation in the case of larger nanostructures, we compare the non-retarded approach with a fully retarded Green's tensor.\cite{paulusAccurateEfficientComputation2000, colasdesfrancsEnhancedLightConfinement2005}
At the example of a larger gold sphere (radius \(r=50\,\)nm) lying on a silica substrate, figure~\ref{fig:retard_non_retard} shows the \(\alpha_{xx}^{EE}\) polarizability tensor element, calculated without (blue line) and with (green line) retardation.
While there is a quantitative deviation in the order of \(\approx5\%\), the qualitative trend is unchanged, whether retardation is included or not.
For comparison we show a spectrum of \(\alpha_{xx}^{EE}\) of an identical sphere but lying on a \(50\,\)nm thick gold film, which is deposited on silica (red dashed line). 
In the latter retardation is again included in the simulation. 
The simulation reveals a strong impact of the plasmonic film on the polarizability of the sphere.

\begin{figure}[t!]
	\centering\includegraphics[width=0.85\columnwidth]{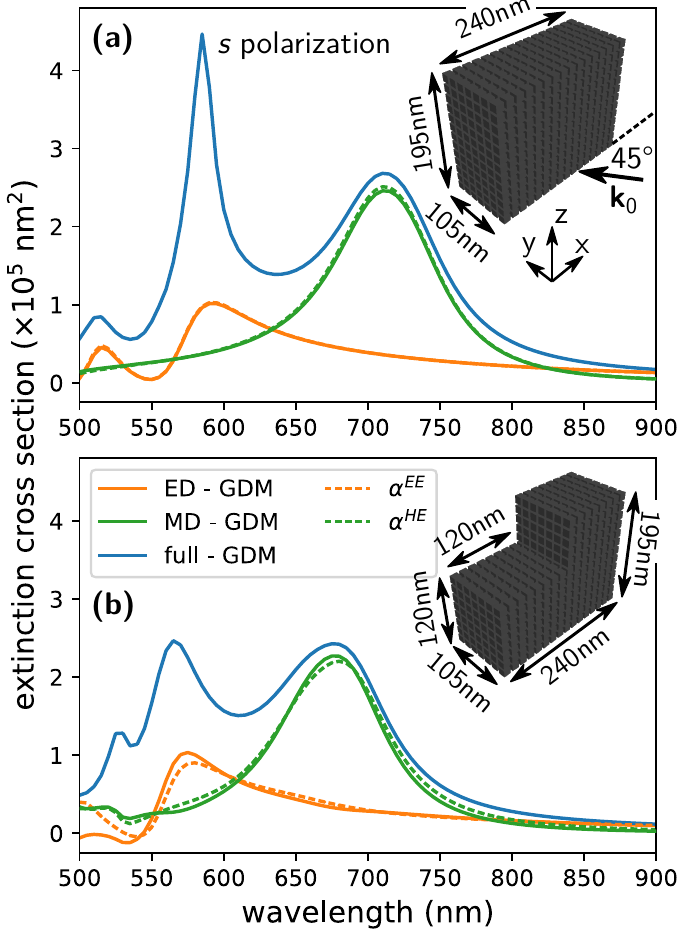}
	\caption{
		An \(s\)-polarized (\(\mathbf{E}_0 \parallel Y\)) plane wave is incident from below at and oblique angle of \(45^{\circ}\) in the \(XZ\) plane. 
		The meso-scale structures are made from a constant index dielectric with \(n=4\), placed in vacuum.
		The figure shows the extinction spectrum calculated via full GDM simulation (solid lines) and obtained from the pseudo-polarizability model (dashed lines) for (a) a ``full'' nano-cuboid and (b) the same cuboid but with on of the top corners removed.
	}
	\label{fig:large_dielectric_approx}
\end{figure}

\begin{figure}[t!]
	\centering\includegraphics[width=\columnwidth]{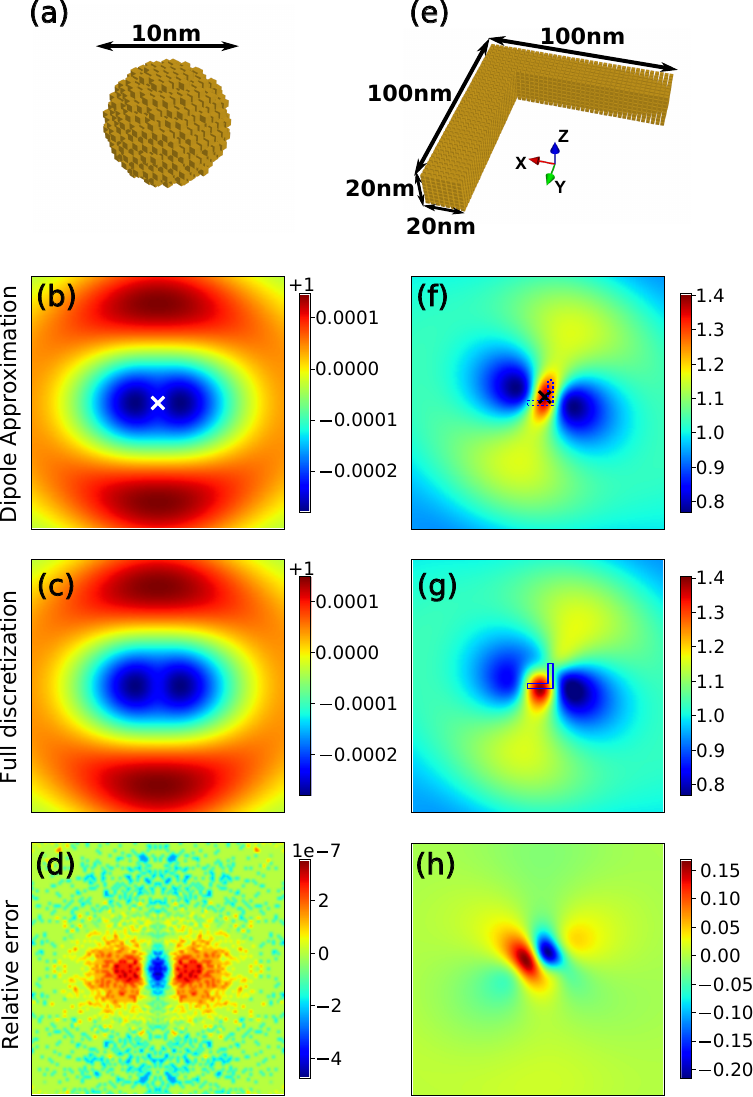}
	\caption{
		Near-field intensity distribution, calculated $100\,$nm above the top-surface of a plasmonic nanostructure for (a-d) a small gold sphere and (e-h) an ``L''-shaped gold nanoparticle.
		(b) and (f) show the near-field calculated using the dipolar effective polarizability approximation.
		(c) and (g) are calculated based on the full GDM simulation.
		(d) and (h) show the relative error of the dipole approximation with respect to the full simulation. 
		(b-c) and (f-g) show the normalized near-field intensity $|\mathbf{E}|^2/|\mathbf{E}_0|^2 $. 
		The incident plane wave (normal incidence) is polarized along \(X\) at a wavelength of \(\lambda_0=690\,\)nm. 
		All colorplots show an area of \(800\times 800\,\)nm$^2$.
	}
	\label{fig:nearfield-breakdown}
\end{figure}

\subsection{Comparison to Mie theory}\label{appendix:mie_theoyr}

We now compare the GDM extracted polarizability of the \(r=50\,\)nm gold sphere to Mie theory and to the full GDM simulation. 
To this end we place the nanosphere in vacuum and illuminate it with a plane wave of linear polarization.
As can be seen in figure~\ref{fig:mie_comparison_goldsphere} the agreement of the extinction efficiency spectra (scattering cross section divided by geometric cross section) is very good with a small quantitative deviation along the slope at the long wavelength side of the plasmon resonance.
We note that the agreement of Mie theory and GDM extracted polarizability and also with the full GDM simulation is better than comparing with the Clausius-Mossotti polarizability, shown in main text figure~\ref{fig:sphere_polarizability}a.

\begin{figure*}[t!]
	\centering\includegraphics[width=\linewidth]{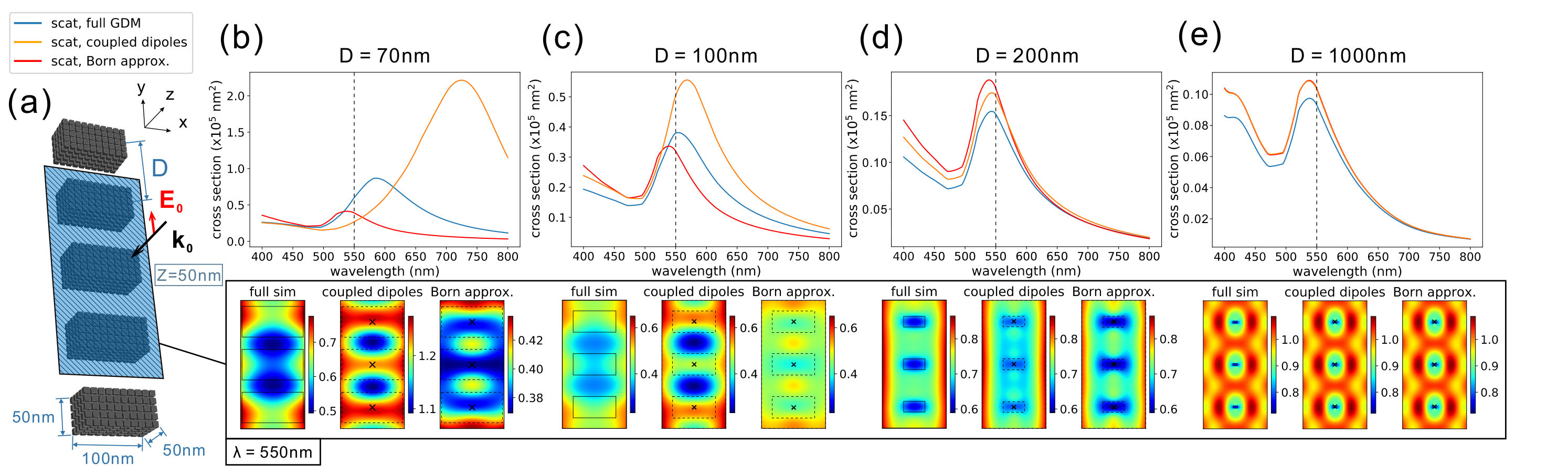}
	\caption{
		Coupling of five gold nanorods of dimensions $100\times 50\times 50\,$nm$^3$ ($X\times Y\times Z$), aligned along the $OY$ axis. 
		As illustrated in (a), the rods are separated by a variable distance $D$ (center-to-center) and are illuminated by a plane wave of linear polarization along $Y$.
		(b-e) For different distances $D$ between the nanorods, comparison of top: scattering spectra and bottom: near-field intensity \(|\mathbf{E}|^2 / |\mathbf{E}_0|^2\) in a plane parallel to $XY$ at a height \(Z=50\,\)nm above the rod's top surface. 
		We compare the full GDM simulation (blue lines, leftmost colormaps) with the effective polarizability model, in which case we either include optical interactions between the dipoles via a self-consistent coupling scheme (orange lines, center colormaps) or we assume that the dipoles are optically isolated, and only interference effects occur (no coupling, corresponding to the Born approximation. Shown as orange lines and in the rightmost colormaps).
		The shown area in the near-field intensity maps is \(3D/2 \times 3D\), the wavelength of the illumination is~\(550\,\)nm. 
		Note that in (b) the near-field maps are not on the same color scale.
	}
	\label{fig:nearfield_coupling_rods}
\end{figure*}

\begin{figure}[t!]
	\centering\includegraphics[width=0.85\columnwidth]{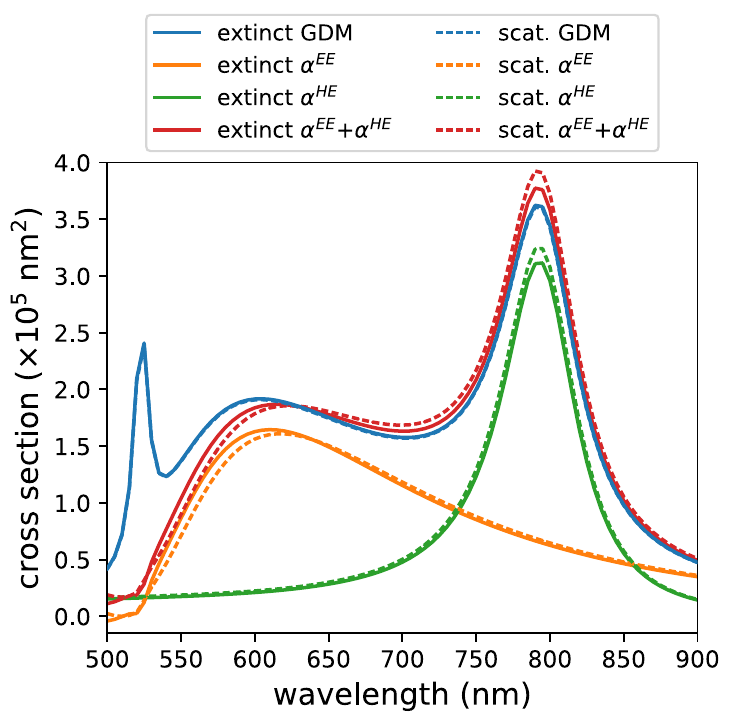}
	\caption{
		Extinction (solid lines) and scattering (dashed lines) cross sections of a dielectric sphere (constant and real refractive index \(n=4\)) with radius \(r=100\,\)nm, placed in vacuum and illuminated by a linear polarized plane wave.
		The sharp resonance around \(525\,\)nm can be attributed to a quadrupole mode.
	}
	\label{fig:scat_vs_ext}
\end{figure}

\subsection{Case of non-symmetric, dielectric structures}\label{appendix:large_dielectric}

Interestingly, our model performs excellent even for quite large structures where \(kR \gtrsim 1\) (see e.g. figure~\ref{fig:interpolation_k}).
We assume that in cases of low symmetry nanostructures, the effective positions of the electric and magnetic dipoles start to move away from the center of mass and can depend on the angle of incidence. 
This may leads to inaccuracies in the superposition approximation which we use for oblique incidence.
We therefore want to assess here how the approach performs on dielectric particles of less symmetric geometries. 

In figure \ref{fig:large_dielectric_approx} we compare a cuboid of side lengths \(240\times 105\times 195\)\,nm\(^3\) \ref{fig:large_dielectric_approx}a and a cuboid of same dimensions but with a missing edge \ref{fig:large_dielectric_approx}b.
In case of the ``bulk'' cuboid in Fig.~\ref{fig:large_dielectric_approx}a, the agreement between dipolar model and full simulation is excellent.
Note that the sharp resonance between \(550\,\)nm and \(600\,\)nm is mainly due to a quadrupolar mode (concerning both geometries), and hence is not described by the dipolar polarizability model.
The increasing asymmetry in case of figure~\ref{fig:large_dielectric_approx}b induces small deviations between polarizability model and full simulation (c.f. solid and dashed green and orange lines in Fig.~\ref{fig:large_dielectric_approx}b), which is probably a results of a non-constant effective position of the effective dipoles under \(X\) and \(Z\) incidence.
The global agreement however remains very good.
We attribute this to the dipole-dominated optical response, even in cases where the nanostructure size is close to the wavelength. The microscopic optical response can be quite complex in such geometries, in particular leading to the formation of optical vortices. Since these vortices effectively act as magnetic dipoles, the \(\alpha^{HE}\) approximation is capable to correctly describe the global response.

\subsection{Breakdown of the dipolar polarizability approximation in the near-field}\label{appendix:near-field}

To a certain extent, the model can be also used to approximate the electromagnetic fields in the vicinity of nanostructures. 
In such case, it is crucial to keep in mind that the polarizabilities describe only a dipolar response, the very proximate near-field can hence not be captured.
To demonstrate the breakdown of the model, we compare in figure~\ref{fig:nearfield-breakdown} the case of a small gold nano-sphere (\ref{fig:nearfield-breakdown}(a-d), diameter of \(10\,\)nm) with the gold ``L''-shaped structure \ref{fig:nearfield-breakdown}(e-h), which was already shown in figure~\ref{fig:complex_structures}(c).
We calculate the electric field intensity in a plane \(100\,\)nm above the nanostructure top-surfaces, and compare the full simulation to the field as given by the dipolar model.

As can be seen, the small sphere (figure~\ref{fig:nearfield-breakdown}(a-d)) behaves almost like a perfect dipole, illustrated by identical near-field maps in figure~\ref{fig:nearfield-breakdown}(b-c), which differ only by a relative error in the order of the machine precision of the $32$ bit floating-point numbers we used in the numerical implementation.
The larger ``L''-shaped nanostructure (figure~\ref{fig:nearfield-breakdown}(e-h)) on the other hand does not exactly behave as a dipole in the near-field region. 
So while the qualitative agreement of the near-field intensity maps in figure~\ref{fig:nearfield-breakdown}(f-g) is still good at a height of $100\,$nm, the peak relative error just above the structure is already as high as around $20\,$\% (dark blue region in Fig.~\ref{fig:nearfield-breakdown}h). 
At even shorter distances the error will drastically increase and the approximation breaks down completely.

\subsection{Near-field coupling between several nanostructures}\label{appendix:nearfield_coupling}

Since the wavevector distribution in the near-field of an assembly of several nanostructures is heterogeneous, the pseudo-polarizability approach would fail to describe such a system of multiple, near-field coupled entities. 
However, in the case of ``static'' effective polarizabilities (without the phase term in Eq.~\eqref{eq:different_alphas}), it is possible to take into account optical interactions between several of such polarizabilities, brought in close vicinity to each other.
This can be done in a self-consistent way using the GDM formalism.\cite{martinGeneralizedFieldPropagator1995, wiechaPyGDMPythonToolkit2018}

Since the model describes scattering at a nanostructure as a dipolar point-scatterer, a certain distance between the individual scatterers is necessary, so that the dipole field is a good approximation for the optical near-field (see also appendix~\ref{appendix:near-field}).
To assess the minimum distance required between several small metallic structures, we show in figure~\ref{fig:nearfield_coupling_rods} simulations of a chain of five gold nano-rods (see illustration in figure~\ref{fig:nearfield_coupling_rods}a), where the distance between the rods is increased successively.
In \ref{fig:nearfield_coupling_rods}(b-e) scattering spectra and near-field intensity maps are compared between the full GDM simulations, the coupled static polarizabilities and finally to the Born approximation, in which scattering of each polarizability is calculated separately, hence only interference effects are taken into account while near-field coupling or multi-scattering events are not considered.
The Born approximation works well only for large spacing values (e.g. $1$\textmu m, shown in \ref{fig:nearfield_coupling_rods}e). 
Re-coupling the static polarizabilities provides a better approximation at shorter distances (e.g. $200$\,nm, as shown in \ref{fig:nearfield_coupling_rods}e).
A very small inter-particle spacing leads then to the breakdown of either approximative model, as can be seen in subfigures~\ref{fig:nearfield_coupling_rods}b and~\ref{fig:nearfield_coupling_rods}c.

Note that the dipole model seems to slightly overestimate the optical cross-sections, which is a systematic observation in agreement with the other plasmonic structures simulations shown throughout this paper.

\subsection{Extinction vs. scattering}\label{appendix:ext_scat}

In order to assess how well the polarizability model is in agreement with energy conservation, we calculate for a lossless dielectric sphere (ref. index \(n=4\), radius \(r=100\,\)nm) the extinction from the optical theorem,\cite{draineDiscreteDipoleApproximationIts1988} and compare it to the scattering cross section.
For a lossless nanostructure, the extinction of the incident light is entirely a result of scattering, hence the cross sections are identical.

The scattering cross section \(C_{\text{scat}}\) can be calculated by re-propagating the effective dipole via the Green's tensor to the far-field, where we integrate the scattered intensity on a sphere of radius \(r_{\text{ff}}=10\,\)\textmu m centred around the nano-structure at \(\mathbf{r}_0\):
\begin{multline}
	C_{\text{scat}} = 
	\Big|\mathbf{E}_0(\mathbf{r}_0,\omega_0)\Big|^{-2} \ 
	\int\limits_{0}^{2\pi}\text{d}\varphi \int\limits_{0}^{\pi}  r_{\text{ff}}^2\ \sin(\vartheta)\ \text{d}\vartheta \\
	\Big|S\big(\mathbf{r_0},\mathbf{r}(\varphi, \vartheta, r_{\text{ff}}),\omega_0\big) \cdot \alpha(\omega_0) \cdot \mathbf{E}_0(\mathbf{r}_0,\omega_0)\Big|^2
	  \, .
\end{multline}
Here \(S\) is the appropriate Green's tensor, depending on the nature of the polarizability \(\alpha\) (electric or magnetic).
\(\mathbf{r}\) is a position on the integration sphere surface, and \(\mathbf{E}_0\) is the complex incident electric field.
The same approach can be used to obtain the scattering section in case of the full GDM simulation by integrating the scattered fields from every meshcell before calculating the field intensity.

In figure~\ref{fig:scat_vs_ext} the full simulation is compared to the superposition of the electric and the magnetic polarizability approximation. 
In case of the full GDM simulation the extinction section through the optical theorem (solid blue line) is in perfect agreement with scattering (dashed blue line).
For both, the electric and the magnetic polarizability, small deviations between extinction and scattering can be observed, meaning that there is some small loss of energy and hence the approximate dipole solution is not perfectly physical. 
The discrepancy is however very small, in particular if one recalls the large size of the sphere (\(kD>1\) over the full spectrum, with here $D=2r=200\,$nm). 
In conclusion, as long as the optical response is dominated by electric and magnetic dipole resonances, our effective pseudo-polarizability approximation offers an excellent performance.

\bibliography{extrac-pola-plasmonic.bbl}

\end{document}